\documentclass[preprint,aps,prb,10pt,showpacs,twocolumn]{revtex4}
\usepackage{epsfig}
\begin{document}

\title{Confined water in the low hydration regime.}

\author{P.~Gallo} \author{M.~Rapinesi} \author{M.~Rovere}\email[Author to whom correspondence should be addressed: ]{rovere@fis.uniroma3.it}
\affiliation{Dipartimento di Fisica, 
Universit\`a ``Roma Tre'', \\ Istituto Nazionale per la Fisica della Materia,
 Unit\`a di Ricerca Roma Tre\\
Via della Vasca Navale 84, 00146 Roma, Italy.}

\begin{abstract}
Molecular dynamics results on water confined in a silica pore 
in the low hydration regime are presented. Strong layering
effects are found due to the hydrophilic character of
the substrate. 
The local properties
of water are studied as function of both temperature and hydration
level. The interaction of the thin films of water
with the silica atoms induces a strong distortion
of the hydrogen bond network. The residence time of
the water molecules is dependent on the distance
from the surface. Its behavior shows a transition
from a brownian to a non-brownian regime 
approaching the substrate in agreement with 
results found in studies of water at contact with globular
proteins.

\end{abstract}

\pacs{61.20.Ja, 61.20.-p, 61.25.-f}

\maketitle
\date{\today}

\section{ Introduction}

Thin films of water confined in restricted geometries
play an important role in many different phenomena in geology, 
chemistry, biology. Important processes taking place
in minerals, plants, biological membranes are determined by
the presence of water confined in nanopores inside the system
or at contact with the surface.~\cite{debenedetti,robinson,israe}    
In biology for instance the enzymatic
activity of globular proteins or the
function of biological membranes are due to the interaction
with water~\cite{debenedetti},
and the decrease of the hydration level may inhibit 
their biological functionality.~\cite{biol}  
It is still difficult to predict accurately 
how the properties of water are modified
when it is at contact with the various substrates or 
confined in different geometries. Nevertheless
recent theoretical and computational work on water confined
in hydrophobic rigid environments indicate 
strong changes in its thermodynamical behavior.~\cite{truskett,meyer}  

A strong reduction of the tetrahedral order with consequent distortion
of the hydrogen bond network has been evidenced by
neutron scattering experiments with isotopic substitution performed
on water confined in Vycor~\cite{mar1,mar2}, by experiments
performed on freezing of water confined in an 
hydrophilic environment of mica~\cite{raviv} and
in water confined in vermiculite 
clay.~\cite{bergman} 
These findings are in agreement with recent computer 
simulation results.~\cite{brovchenko,jmoliq,gr-jcp} 
Particularly relevant are the observations in
experiments and computer simulation
of the formation of distinct layer of water with
different structural and dynamical properties: an interfacial layer
close to the substrate, where water molecules show a slow dynamics, and
a layer far from the surface with almost bulk 
like properties.~\cite{melni,kremer,barut,mckenna,noi-epl,noi-prl,noi-jcp}   
The layering effect is expected to be relevant in many phenomena
associated with confined water. There are experimental evidences
of a slow relaxation of water at contact with 
proteins~\cite{doster,bizzarri1,bizzarri2,bizzarri3,bizzarri4},
while signatures of a similar behavior has been found in water 
confined in Vycor for low hydrations.~\cite{venturini}

As long as  
a detailed description of the 
microscopic properties  of water at interface with 
complex materials like biological macromolecules
is difficult to be obtained it could be
appropriate to use model systems with well defined
geometry and interaction between water and substrate. 
We present here the results obtained by a molecular dynamics (MD)
study of a model of water confined in Vycor at low hydration levels.
By computer
simulation we can freely vary temperature and hydration level
and make an accurate study of the changes induced
in the properties of water by the confinement.
This work is part of a detailed study of the structural
and dynamical properties 
of confined water.~\cite{jmoliq,gr-jcp,noi-epl,noi-prl,noi-jcp}

The low hydration regime is somehow more important to be explored
at least for two main reasons:
(i) the effect of the substrate on the confined water is
expected to be enhanced; (ii) the experimental observations
are the result of an averaging over different water layers and
an accurate analysis of the low hydration regime is the only
way to gain some insight on the water-substrate interaction.    

An accurate layer analysis enables us to examine
the modifications of the main structural properties
of confined water and to make connection with recent 
experimental and computer simulation results which
concern water at contact with proteins.

In Sec. II we briefly describe our simulation method, 
in Sec. III, IV and V we present the main results of our
calculation. Sec. VI is devoted to conclusions.

\section{ Computer simulation of confined water}

The computer simulation is performed in the microcanonical
ensemble on 
water molecules described by the simple point charge/extended 
(SPC/E) site model potential 
inserted in a cylindrical cavity of radius $20$~\AA.
The cavity has been carved in a cubic cell of silica glass obtained
by computer simulation, as described in previous 
work.~\cite{jmoliq,gr-jcp,noi-jcp}
The system is modeled to represent water confined in Vycor, 
the internal surface of the cylindrical cell is corrugated and composed
by silicon atoms (Si), bridging oxygens (BO) which are bonded to
two silicons and non-bridging oxygens (NBO) connected only to one 
silicon. The NBO are saturated with acidic hydrogens (AH). 
The interaction of water with the substrate is 
modeled by assigning different charges to the sites representing
the atoms of Vycor. A Lennard-Jones interaction is assumed
between the oxygens of water and the BO and NBO of the confining
system.~\cite{jmoliq} Periodic boundary conditions
are applied along the axis of the cylinder assumed as the
$z$ direction. The motion is confined
in the $xy$ plane, where the distance from the axis
is defined in terms of the radius $R=\sqrt{x^2+y^2}$.
During the simulation the substrate is kept rigid.

\begin{figure}
\centering\epsfig{file=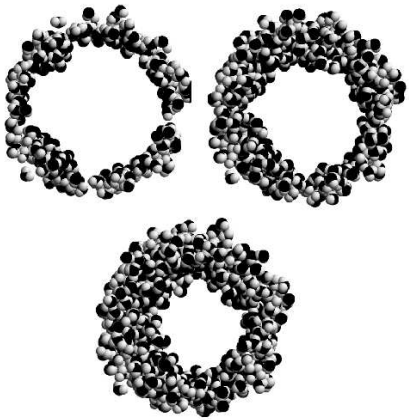,width=0.8\linewidth}
\caption{snapshots of the configuration of water molecules
at room temperature: from above and from the left  $N_W=500$, $N_W=1000$,
$N_W=1500$.
}
\protect\label{fig:snap}
\end{figure} 
The density
of the full hydration in the experiments~\cite{mar1} $0.0297 $~\AA$^{-3}$ 
corresponds in our system to roughly $N_W=2600$ molecules. We performed
our simulation at three different hydration levels $N_W=500$ 
which corresponds to $19$~\% of hydration,
$N_W=1000$ ($38$~\%) and $N_W=1500$ ($58$~\%).   
In Fig.~\ref{fig:snap} the snapshots of the configurations at 
the three different
hydration levels show the hydrophilic character of the substrate.

\begin{figure}
\centering\epsfig{file=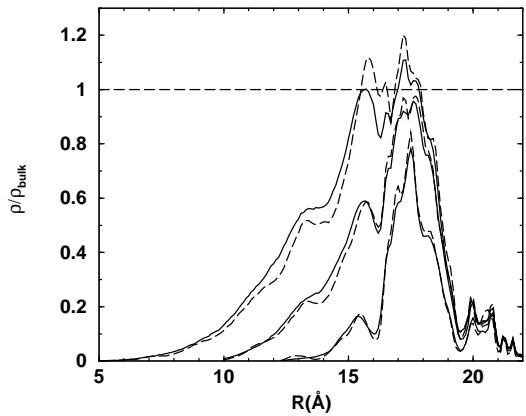,width=0.8\linewidth}
\caption{
Density profiles of the confined water along the pore radius
for different hydration levels: from below 
$N_W=500$, $N_W=1000$, $N_W=1500$ at two different
temperatures: $T=300$~K (solid line), $T=220$~K (dotted line).
The density is normalized to the value for water at ambient temperature 
$\rho_{bulk}=0.0334$~\AA$^{-3}$.}
\protect\label{fig:profili}
\end{figure} 
\section{ Density profiles}

In the density profiles reported in Fig.~\ref{fig:profili} 
it is evident
for all the hydrations the formation of a well defined first layer of
molecules close to the surface, while a second layer grows up 
with the hydration level. Few molecules penetrate and
are trapped inside the substrate ($R > 20$~\AA). 
We notice that at the highest hydration
($N_W=1500$) the density in the double layer structure increases
above the value of density of water at ambient conditions. The decrease
of temperature has very little effect on the density profiles, apart
from the sharpening of the second layer at the hydration $N_W=1500$. 

\begin{figure}[t]
\centering\epsfig{file=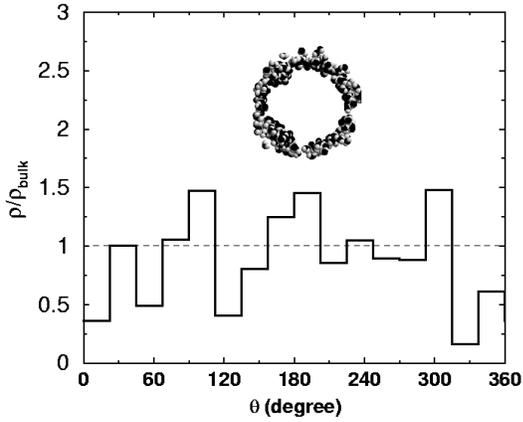,width=0.8\linewidth}
\caption{ 
Instantaneous 
density profile for $N_W=500$ at room temperature in the $xy$ plane 
along a circumference 
of radius $R=17.5$~\AA\ corresponding to a maximum in the 
average density profile
at the same hydration (see Fig.~\ref{fig:profili}). The density profile
is obtained from the snapshot in Fig.~\ref{fig:snap} and shows the
arrangement of the molecules in clusters.}
\protect\label{fig:circo}
\end{figure}

It is clear from Fig.~\ref{fig:snap} that the arrangement of the molecules
close to the Vycor surface is not uniform and the formation of clusters
of different sizes is observed, this is even more evident 
in Fig.~\ref{fig:circo} where we report
an histogram of the distribution of the molecules in the $xy$ plane along 
a circumference at distance $R=17.5$~\AA\  where the density profile
reaches the maximum. 
The cluster sizes fluctuate up to hundred molecules with
an average value of $40$ molecules, this is in agreement with
neutron diffraction experiments~\cite{agamalian} where clusters
of the same size are observed at similar value of the hydration.

\section{Layer analysis of the hydrogen bond}

\begin{figure}[t]
\centering\epsfig{file=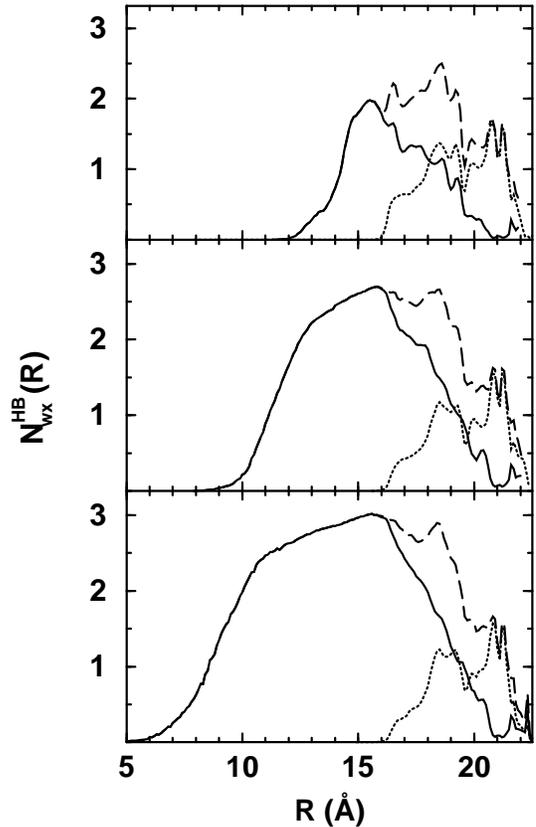,width=0.8\linewidth}
\caption{ Variation of the average number of hydrogen bonds per 
molecule along the pore radius at $T=300$~K for $N_W=500$ (top),
$N_W=1000$ (middle) and $N_W=1500$ (bottom) at room
temperature. The solid line is the water-water $N^{HB}_{ww}(R)$,
the dotted line is the water-Vycor $N^{HB}_{wv}(R)$ and the long dashed
line represents the total $N^{HB}_{tot}(R)$. 
}
\protect\label{fig:legami}
\end{figure}

Figure~\ref{fig:legami} shows the average number of hydrogen
bond(HB) per molecule
along the pore radius for the different hydration levels.
We assume that two neighbors water molecules are 
hydrogen bonded if the $H-O \cdots O$ angle is
less than $30^o$.~\cite{jedlov}
Two water molecules are considered neighbors  
if their O-O separation is less than $3.35$~\AA.
The number of water-water HB $N^{HB}_{ww}(R)$ 
decreases approaching the surface, while inside the double layer structure
it grows up the number of HB of the water with
the Vycor surface $N^{HB}_{wv}(R)$. The water molecules form preferentially 
hydrogen bonds with the bridging oxygens. 
The arrangement of the
clusters close to the Vycor surface is mainly determined by this type of
bonds. 
The number of HB between water
molecules and the acidic H atoms is negligible on the scale of the figure.    
The $N^{HB}_{wv}(R)$ curves do not show substantial differences as function
of hydration level of the pore.

\begin{figure}[t]
\centering\epsfig{file=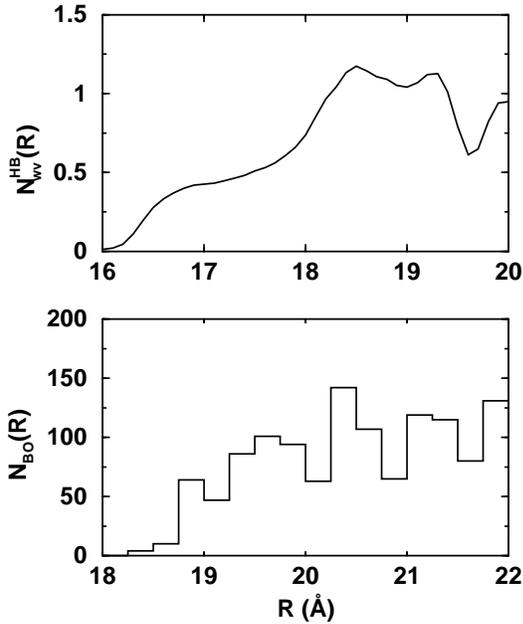,width=0.8\linewidth}
\caption{The profile of the average number of water-Vycor hydrogen bonds
per molecule along the pore radius (top) compared with a
profile of the number of bridging oxygens (bottom) on the Vycor
substrate ($N_W=1000$ and $T=300$~K).
}
\protect\label{fig:legamiwv}
\end{figure}

In Fig.~\ref{fig:legamiwv} the radial profile of the water-Vycor bonds
is compared with the density profile of the bridging oxygens at the
surface; by considering that the bond length is approximately
$2$~\AA\ there is a clear correspondence between the two profiles.
It is evident that the bridging oxygens
substitute as acceptors the water oxygens approaching the surface 
since the total number of HB 
$N^{HB}_{tot}(R)=N^{HB}_{ww}(R)+N^{HB}_{wv}(R)$ , 
also reported in Fig.~\ref{fig:legami}, remains
practically constant. 
The maximum number of water-water HB grows with increasing
hydration level since more water molecules become available
for the formation of bonds. The maximum is located around $R=16$~\AA\
where the density profile shows a minimum (see Fig.~\ref{fig:profili}),
in agreement with the fact that the hydrogen bonds increase with
decreasing density at room temperature and upon supercooling.~\cite{geiger}

At lower temperatures the number of water-water and
water-Vycor hydrogen bonds increases but the profiles are
similar to those shown in Fig.~\ref{fig:legami} at $T=300$~K.
 
\begin{figure}[h]
\centering\epsfig{file=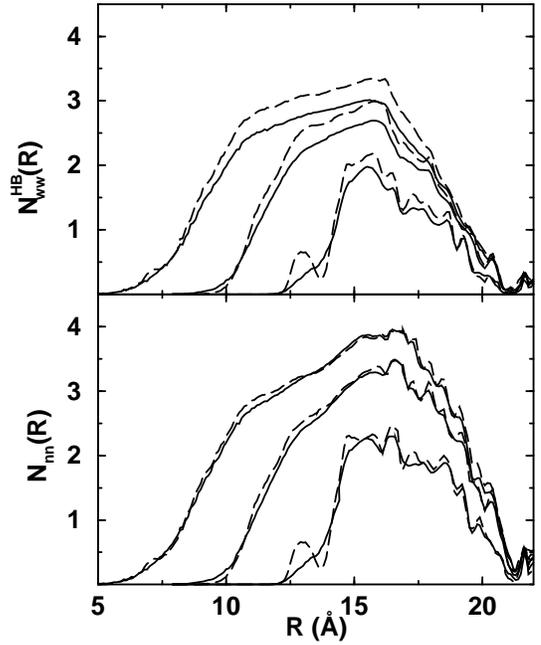,width=0.8\linewidth}
\caption{ Comparison of the water-water hydrogen bonds per 
molecule $N^{HB}_{ww}(R)$ (top) and the average number of
nearest-neighbor water molecules $N_{nn}(R)$ (bottom) along 
the pore radius for the hydration level $N_W=1500$,
$N_W=1000$ and $N_W=500$ in descending order at  
temperatures $T=300$~K (solid),
$T=220$~K (long-dashed).
}
\protect\label{fig:legamiww}
\end{figure}

\begin{figure}
\centering\epsfig{file=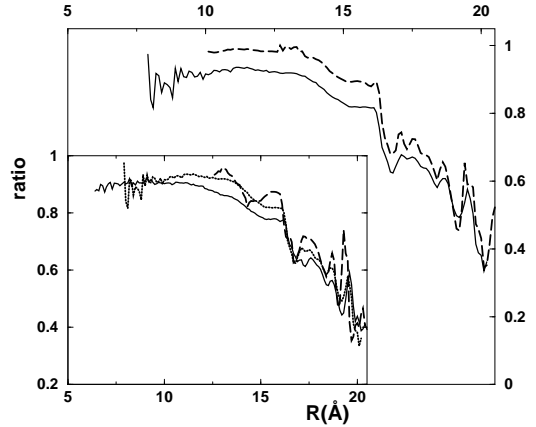,width=0.8\linewidth}
\caption{ Ratio $N^{HB}_{ww}(R)/N_{nn}(R)$ along 
the pore radius for  $N_W=1000$ at temperatures $T=300$~K (solid),
$T=220$~K (long-dashed). In the inset  $N^{HB}_{ww}(R)/N_{nn}(R)$
is reported at $T=300$~K for the hydrations  
$N_W=1500$ (solid line),
$N_W=1000$ (dotted line) and $N_W=500$ (long dashed line).
}
\protect\label{fig:ratiowwnn}
\end{figure}

In Fig.~\ref{fig:legamiww} we report the radial profiles of the 
average number of water-water hydrogen bonds compared  
with the profiles of the average number of neighbors $N_{nn}$. 
The ratio between the two quantities $N^{HB}_{ww}(R) / N_{nn}(R)$ 
estimates the number of molecules forming 
hydrogen bonds between the neighbors. It can be 
considered as a measure of the rigidity of the hydrogen bond
network and it is reported in Fig.~\ref{fig:ratiowwnn} 
for selected values of the hydration and temperature.
The ratio is smaller close to the surface ($R>16$~\AA) where
the hydrogen bonds are made mainly with the Vycor surface.
Below $16$~\AA\ each molecule is able to make hydrogen 
bonds with almost all its neighbors.
 
The network becomes more rigid on lowering the temperature 
(see Fig.~\ref{fig:ratiowwnn}) since
the number of hydrogen bonds increases while
the number of neighbors is almost insensible to the
variation of temperature, like in the density profiles 
in Fig.~\ref{fig:profili}. This might be possibly related to the
recently proposed fragile to strong transition for strongly confined
water.~\cite{bergman} 

We can conclude that
the water molecules rearrange themselves at decreasing temperature in order
to maximize the number of hydrogen bonds. 
We observe moreover that the rigidity increases at 
decreasing hydration (see inset of Fig.~\ref{fig:ratiowwnn}) 
with some important
consequences for the residence times, as discussed later.

\begin{figure}[t]
\centering\epsfig{file=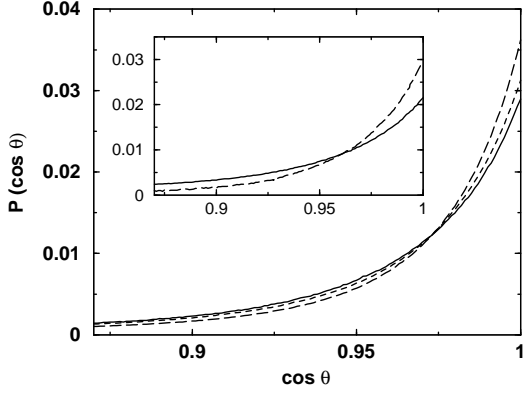,width=0.8\linewidth}
\caption{Hydrogen bond angle distribution function, $P(\cos \theta)$ 
(see text for definition) for $N_W=1000$ and different temperature: 
$T=300$~K (solid),$T=260$~K (dashed), $T=220$~K (long dashed). In the
inset for $N_W=500$, layer analysis of
the distribution at T=300, $0 <R< 16$~\AA\ (long dashed), 
$16 <R< 20$~\AA\ (solid).
}
\protect\label{fig:angolo1}
\end{figure}

We investigated how the short range order present in 
bulk liquid water is modified upon confinement by looking at
the distribution of the hydrogen bond angles.
Fig.~\ref{fig:angolo1} shows the distribution of $cos(\theta)$, where 
$\theta$ is the angle between the intramolecular bond O-H and the 
intermolecular O$\cdots$O vector, for $N_W=1000$ and different 
temperatures. 
The water-water linear hydrogen bonds are still 
favored like in bulk water, there is however a more pronounced tail 
toward large angle with respect to bulk water. By lowering the temperature 
the distribution peak becomes sharper indicating an increase in
the rigidity of the bonds. The distortion in the angular distribution
with respect to the bulk
is due to the water molecules close to the pore surface, as clearly appears
in the inset of Fig.~\ref{fig:angolo1}, where for $N_W=500$ the 
separated contributions
from the layer $0 <R< 16$~\AA\ and $16 <R< 20$~\AA\ are shown.   
For the molecules close to the surface the peak of the angular distribution 
is depressed while there is a large increase of the tail at high angle,
the angular distribution of the molecules closer to the interior of
the pore is similar to the one of the bulk water.

The distortion of the hydrogen bonds is confirmed by experimental
studies on water confined in Vycor~\cite{mar1,mar2} and at contact with
other surfaces.~\cite{raviv} 

\begin{figure}[h]
\centering\epsfig{file=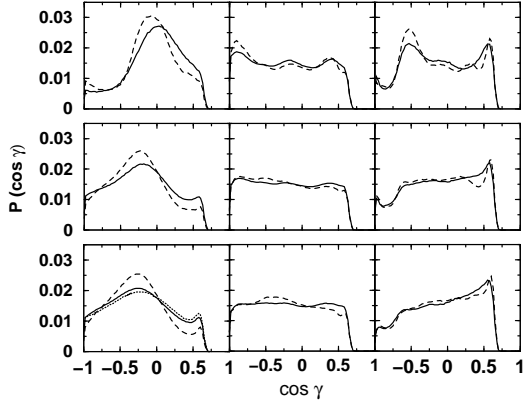,width=0.8\linewidth}
\caption{Layer analysis of the angular distributions of the angle between the
oxygens of three nn water molecules, shown for $N_W=500$ (top),
$N_W=1000$ (middle), $N=1500$ (bottom) at $T=300$~K (solid line) and
$T=220$~K (dashed line). For each hydration the figure shows
the angular distribution for the layer $0 <R< 16$~\AA\ on the left,
the layer  $16 <R< 18$~\AA\ in the center and the layer  $18 <R< 20$~\AA\
closest to the substrate on the right. For comparison 
the angular distribution for ambient water is shown as a dotted line
in the first inset at bottom on the left.
}
\protect\label{fig:angolo2}
\end{figure}

We studied by layer analysis the local arrangement of the 
nearest neighbor molecules by looking to the distribution
of the angle $\gamma$ between the 
two vectors joining the oxygen atom of a water molecule with the oxygen
atoms of the two closest water neighbors. This distribution in the
computer simulation of bulk
water at ambient condition has a well defined peak 
around $\gamma \approx 109^o$ ($\cos(\gamma) \approx -0.326$),
this peak is a signature of the tetrahedral order
present in liquid water at short range distance. There is
also a secondary peak at $\gamma \approx 54^0$  
corresponding to interstitial neighbor molecules. These findings
are in agreement with experimental results on water at ambient
conditions.\cite{jedlov}
In Fig.~\ref{fig:angolo2} we report a layer analysis of the distribution 
function of $\cos(\gamma)$ of water in Vycor for different hydrations.   
The molecules in the layer $0 <R< 16$~\AA\ 
for $N_W=1500$ and $N_W=1000$ show
a distribution very similar to what found in the simulation of bulk water,
the distributions become sharper at lower temperature.
For $N_W=500$ the peak shifts toward $90^o$, the value at which
the maximum is located for $T=220$~K. This behavior for the lowest
hydration is in agreement with experimental results 
indicating that in confined water freezing in a cubic crystalline phase
could be favored.~\cite{rrgrenoble}

As a matter of fact a good short range order
is present for all the hydrations
in the layer $0 <R< 16$~\AA\ where the water molecules are 
able to form hydrogen bonds with almost all their 
neighbors as discussed above (see Fig.~\ref{fig:legamiww}).

In the layer closest to the Vycor surface 
$16 <R< 18$~\AA\ the tetrahedral
arrangement is completely lost due to the formation of hydrogen
bonds between water molecules and the substrate. The temperature
has very little effect on these distributions.
In the layer $18 <R< 20$~\AA\ the specific shape 
of the distribution is not very
significant and indicates the strong distortion of the
hydrogen bonds formed by the molecules found in the inner surface
of the pore.

\section{Residence time}

\begin{figure}[h]
\centering\epsfig{file=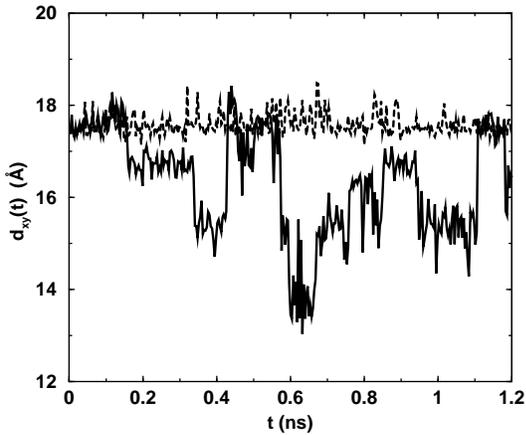,width=0.8\linewidth}
\caption{Trajectory of the center of mass of a molecule
in the $xy$ plane $d_{xy}(t)=\sqrt{x^2(t)+y^2(t)}$ for
$N_W=500$ and temperature $T=300$~K (black line), 
$T=220$~K (dashed line). 
}
\protect\label{fig:dist}
\end{figure}

The dynamics of the molecules of confined water is strongly determined
by the interaction with the substrate. In Fig.~\ref{fig:dist} 
is reported the trajectory of the center of mass of a molecule     
which resides at the beginning in the first layer at roughly $2$~\AA\
from the surface for the case $N_W=500$. At ambient temperature
the molecule oscillates mainly between the two layers and
persists in a position for fractions of nanoseconds, at lower
temperature the molecule oscillates for all the time explored
around the same position and does not go out from its initial layer.

\begin{figure}[h]
\centering\epsfig{file=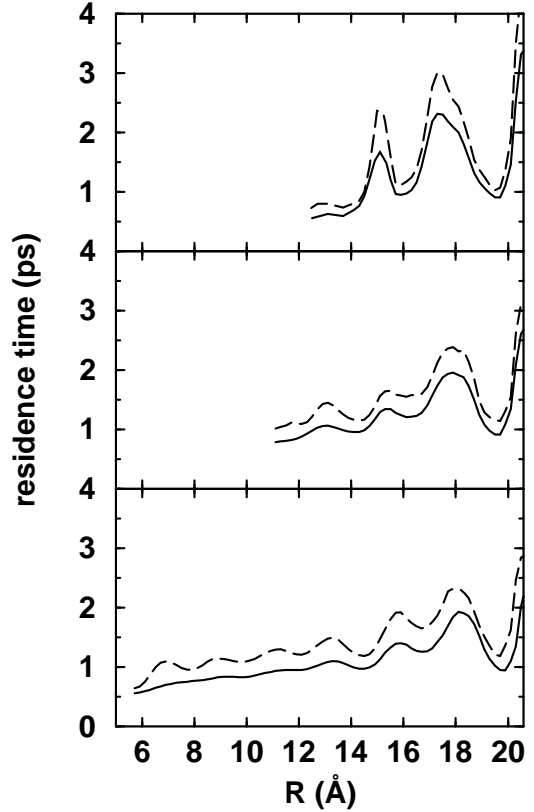,width=0.8\linewidth}
\caption{Average residence time of the water molecules along
the pore radius for $T=300$~K (solid line) 
and $T=220$~K (long dashed) at different hydrations:
$N_W=500$ (top), $N_W=1000$ (middle), $N_W=1500$ (bottom).
}
\protect\label{fig:tempimedi}
\end{figure}

Fig.~\ref{fig:tempimedi} shows how the average residence time 
of the water molecules changes along the pore radius
for different hydrations and temperatures. 
The residence time for each region is calculated as
\begin{equation}
t_{res}= \sum_{i=1}^{N_W} \sum_{k=1}^{q_i}
\Delta t_{ik}/\sum_{i=1}^{N_W} q_i
\end{equation}
where $\Delta t_{ik}$ is the fraction of time that
the molecule $i$ spent during the $k$-th visit in
the region. 
For all hydrations
oscillations are present which are particularly marked for $N_W=500$.
The residence time reaches the maximum values at the surface.
The structure of the residence time profiles around the two main peaks  
is reminiscent of the double layer structure of the density profiles
(see Fig.~\ref{fig:profili}).
Apart for the layer attached to the Vycor surface
the molecules have longer residence time in the regions where
the density profiles have local maximal values, while
the residence time minimum is found roughly at the minimum of
the density distribution between the two layers. 
The water molecules stay for a shorter time in the range between the
two layers where the maximum of the water-water HB profile is located. 
So in the region of minimum density the HB formed are not very stable. 

The maxima of the residence time are enhanced
by decreasing the hydration.
As discussed in Sec. IV
this effect can be explained in terms of an increase
of the rigidity of the hydrogen bond network at decreasing
hydration.

Close to the substrate where an 
elevated number of water-Vycor hydrogen
bonds $N^{HB}_{wv}$ are present a comparison of 
the average residence time in the zone $16 <R< 20$~\AA\
with $N^{HB}_{wv}$ shows that the two quantities 
are proportional in the range of temperature explored  
\begin{equation}
t_{res}(T) = {\cal{C}}_N N^{HB}_{wv}(T) \label{eq:tres}
\end{equation}
where the constant ${\cal{C}}_N$ increases at decreasing hydration level
going from $10$ for $N_W=1500$ to $20$ for $N_W=500$.

\begin{figure}[h]
\centering\epsfig{file=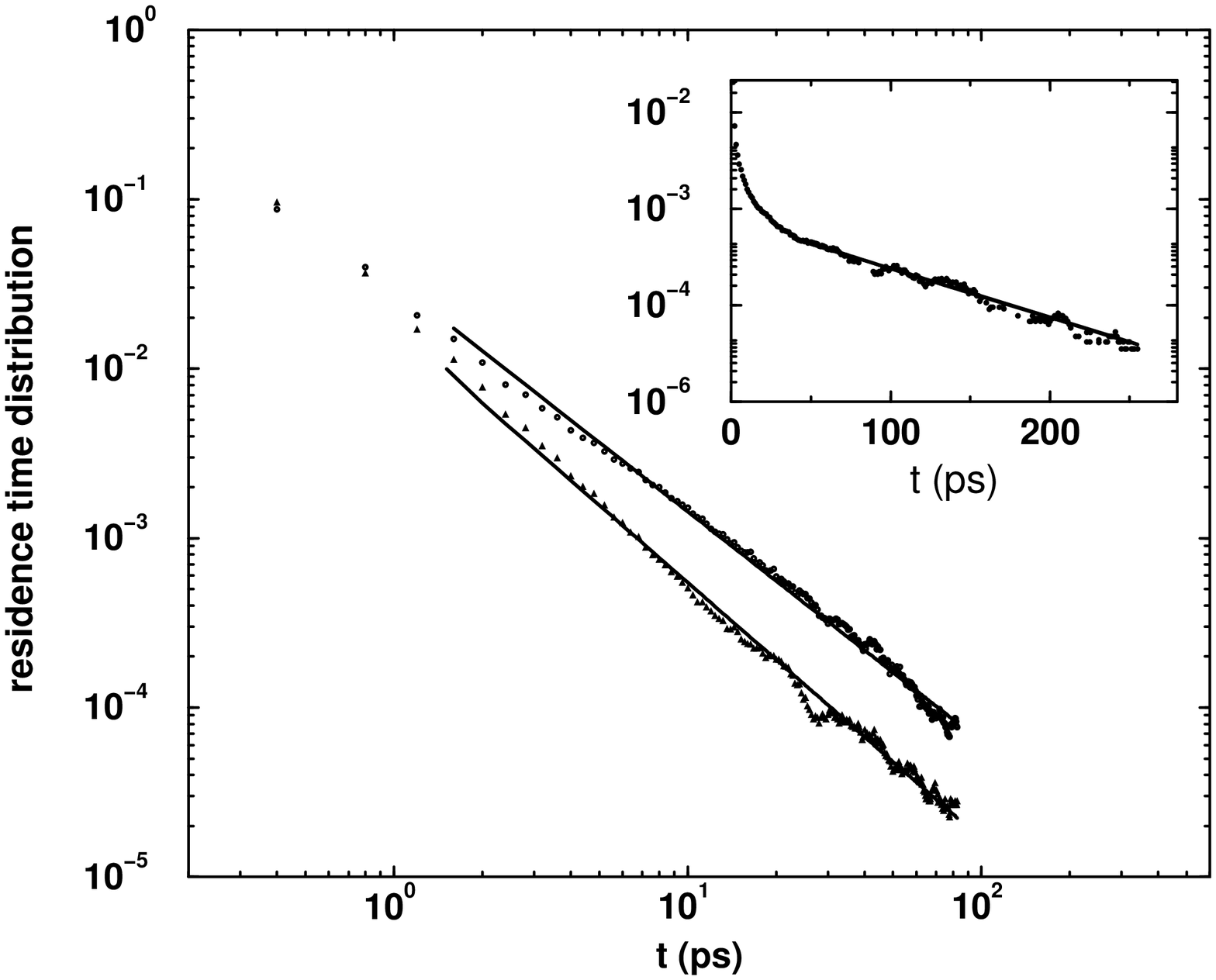,width=0.8\linewidth}
\caption{Residence time distribution at room temperature for $N_W=1500$ 
calculated for different layers: $16< R <20$~\AA\ (triangles)
and bottom curve,
$14< R <16$~\AA\ (circles) and
top curve, in both cases
the solid line is the power law fit (see Eq.~\ref{eq:powlaw}). In the inset is 
reported the result for the layer $0< R <16$~\AA\,
the solid line is the fit to the exponential law of Eq.~(\ref{eq:explaw}).
}
\protect\label{fig:rtd}
\end{figure}
The distributions of the residence times in the different layers show
peculiar features, as illustrated in Fig.~\ref{fig:rtd}
for $N_W=1500$ at room temperature. 
The distributions ${\cal{P}}(t_{res})$ for the layers more close to
the surface are well described by a power law 
\begin{equation}
{\cal{P}}(t) = a t^{-\mu}  \label{eq:powlaw}
\end{equation}
where for the molecules in layer $16< R <20$~\AA\ 
$\mu= 1.51$, while by considering the layer
$14< R <20$~\AA\ $\mu=1.36$.
This behavior is a signature of a non brownian motion of the
particles and it is observed at all the hydration levels 
explored in this work.
Non brownian diffusion has been found recently in neutron
scattering experiment and MD simulation on water at contact
with a globular protein~\cite{bizzarri3} where the
distribution of the residence times of the water molecules in 
the layers close to the protein surface
is fitted to Eq.~(\ref{eq:powlaw}) with exponents $\mu$ in fairly
good agreement with the exponents found in the present work.

For water molecules far from the protein surface it is found~\cite{bizzarri3}
that the brownian regime is established with
an exponential decays of the residence time distribution. 
In the inset of Fig.~\ref{fig:rtd} the residence time distribution 
is reported for the internal layer 
$0< R <16$~\AA\ of our system, also in our case the decay at long time
can be fitted with an exponential
 \begin{equation}
{\cal{P}}(t) = b e^{-Bt}  \label{eq:explaw}
\end{equation}

\section{Summary and Conclusions}

The structural properties of water confined in a cavity 
representing the average properties of the pores of Vycor glass 
have been studied by computer simulation. Water molecules
are attracted by the hydrophilic surface and form a double
layer structure of $4 \div 5$~\AA. In this region it is found
that the local properties of water are 
strongly dependent on the hydration level and the distance
from the substrate. 

The hydrogen bond network is deformed with respect to bulk water,
in agreement with experimental results.~\cite{mar1,mar2,raviv}
At a distance greater than $4$~\AA\ from the surface
for the highest hydrations investigated ($N_W=1500$ and $N_W=1000$) 
the water molecules are arranged in a  
local tetrahedral order similar to bulk water, while
approaching the substrate the water-Vycor hydrogen bonds substitute
the water-water hydrogen bonds and 
the tetrahedral order is completely lost.    
At the lowest studied hydration ($N_W=500$) the local order
far from the surface is compatible with nucleation 
of a cubic crystalline phase at low temperature. 
Experimental evidences for nucleation of cubic ice has been indeed found
for water under confinement.~\cite{rrgrenoble}

At a given hydration the average number of water-water HB increases 
with decreasing temperature while the number of nearest neighbors remains
constant, this implies that the hydrogen bond
network becomes more rigid upon supercooling. The effect is
accentuated by decreasing the hydration. This findings 
are consistent with
the recently proposed fragile to strong transition for
strongly confined water.~\cite{bergman} 

The analysis of the residence time of the water molecules 
along the pore radius evidenced connections between
this quantity and the density profiles, the water molecules
persist for longer time in the regions of the maxima of
the double layer structure and very close to the Vycor surface.

The residence time distributions show an anomalous non brownian
behavior in the layers close to the substrate with a power law decay for long
time. On the contrary the distributions decay exponentially 
in the layers far from the surface. Experimental and computer simulation 
studies on water
at contact with a globular protein~\cite{bizzarri3}
show a similar behavior. This agreement
indicates that there are
useful analogies between the behavior of water confined in Vycor 
and the behavior of water at contact with proteins.
This is confirmed by a more refined analysis of
the dynamical properties of confined water in the low
hydration regime.~\cite{lowhdyn}

\section {Acknowledgments}

We thank M. A. Ricci for critical reading of the manuscript.



\end{document}